# The evolution of cooperation by social exclusion

Tatsuya Sasaki[a,1] and Satoshi Uchida[b]

[a]Evolution and Ecology Program, International Institute for Applied Systems Analysis, Schlossplatz 1, 2631 Laxenburg, Austria

[b]Research Center, RINRI Institute, Misaki-cho 3-1-10, Chiyoda-ku, 101-8385 Tokyo, Japan

[1]To whom correspondence should be addressed.

E-mail: sasakit@iiasa.ac.at

6 December 2012

Preprint version 2.0

**Summary:** The exclusion of freeriders from common privileges or public acceptance is widely found in the real world. Current models on the evolution of cooperation with incentives mostly assume peer sanctioning, whereby a punisher imposes penalties on freeriders at a cost to itself. It is well known that such costly punishment has two substantial difficulties. First, a rare punishing cooperator barely subverts the asocial society of freeriders, and second, natural selection often eliminates punishing cooperators in the presence of non-punishing cooperators (namely, "second-order" freeriders). We present a game-theoretical model of social exclusion in which a punishing cooperator can exclude freeriders from benefit sharing. We show that such social exclusion can overcome the above-mentioned difficulties even if it is costly and stochastic. The results do not require a genetic relationship, repeated interaction, reputation, or group selection. Instead, only a limited number of freeriders are required to prevent the second-order freeriders from eroding the social immune system.

**1. Introduction**

We frequently engage in voluntary joint enterprises with nonrelatives, activities that are fundamental to society. The evolution of cooperative behaviors is an important issue because without any supporting mechanism [1], natural selection often favours those that contribute less at the expense of those that contribute more. A minimal situation could easily cause the ruin of a commune of cooperators, namely, the "tragedy of the commons" [2]. Here we consider different types of punishment, such as a monetary fine (e.g., [3–7]) and ostracism (e.g., [8–11]), for the evolution of cooperation. Punishment can reduce the expected payoff for the opponent, and subsequently, change natural selection preferences, to encourage additional contributions to communal efforts [12]. Our model looks at this situation, because "very little work has addressed questions about the form that punishment is likely to take in reality and about the relative efficacy of different types of punishment" [13].

Here, we choose to focus on social exclusion, which is a common and powerful tool to penalise deviators in human societies, and includes behaviors such as eviction, shunning and ignoring [14–16]. For self-sustaining human systems, indeed, the ability to distinguish among individuals and clarify who should participate in the sharing of communal benefits is crucial and expected (of its members) [17]. A specific example is found in the case of traffic violators who are punished, often strictly by suspending or revoking their driver license for public roads. Among non-humans, shunning through partner switching is a common mechanism for inequity aversion and cooperation enforcement [13,18,19]. Experimental studies have shown, for instance, that chimpanzees can use a mechanism to exclude less cooperative partners from potential collaborations [20], or that reef fish will terminate interaction with cleaner fish that cheat by eating the host's mucus rather than parasites [21].

In joint enterprises, by excluding freeriders from benefit sharing, the punishers can naturally benefit, because such exclusion often decreases the number of beneficiaries, with little effect on the total benefit. Consider the example of the division of a pie provided by some volunteers to a group. If a person is one of the volunteers, it may be justifiable in terms of fairness to suggest or even force freeriders to refrain from sharing in the pie. Although excluding freeriders can be stressful, it increases the share of the pie for the contributors, including the person who performs the actual exclusion. If the situation calls for it, the excluded freerider's share of the group benefits may separately be redistributed among the remaining members in the group [22,23]. Therefore, in either case, the excluded member will obtain nothing from the joint enterprise and the exclusion causes immediate increases in the payoff for the punisher and also the other remaining members in the group.

This is a "self-serving" form of punishment [13,18]. It is of importance that if the cost of excluding is smaller than the reallocated benefit, social exclusion can provide immediate net benefits even to the punisher. This can potentially motivate the group members to contribute to the exclusion of freeriders, however, our understanding of how cooperation unfolds through social exclusion is still "uncharted territory" [24].

Most game-theoretical works on cooperation with punishment have focused on other forms of punishment, for example, costly punishment that reduces the payoffs of both the punishers and those who are punished. As is well known, costly punishment poses fundamental puzzles with regard to its emergence and maintenance. First of all, costly punishment is unlikely to emerge in a sea of freeriders, in which almost all freeriders are unaffected, and a rare punisher would have to decrease in its payoff through punishing the left and right [18,25–27]. Moreover, although initially prevalent, punishers can stabilise cooperation, while non-punishing cooperators (so-called "second-order freeriders") can undermine full cooperation once it is established [3,13,17,24,29].

In terms of self-serving punishments, however, we have only started to confront the puzzles that emerge in these scenarios. We ask here, what happens if social exclusion is applied?: that is, do players move toward excluding others?, and can freeriders be eliminated? Or, will others in the group resist? Our main contribution is to provide a detailed comparative analysis for social exclusion and costly punishment, two different types of punishment, from the viewpoint of their emergence and maintenance. With the self-serving function, social exclusion is predicted to more easily emerge and be maintained than costly punishment.

Few theoretical works have investigated the conditions under which cooperation can evolve by the exclusion of freeriders. Our model requires no additional modules, such as a genetic relationship, repeated games, reputation, or group selection. Considering these modules is imperative for understanding the evolution of cooperation in realistic settings. In fact, these modules may have already been incorporated in earlier game-theoretical models that included the exclusion of freeriders [30–32], but we are interested in first looking at the most minimal of situations to get at the core relative efficacy of costly punishment versus social exclusion.

**2. Game-theoretical model and analysis**

To describe these punishment schemes in detail, we begin with standard public good games with a group size of $n \geq 2$ (e.g., [26,33,34]) in an infinitely large, well-mixed population of players. We specifically apply a replicator system [35] for the dynamic analysis, as based on preferentially imitating strategies of the more successful individuals. In the game, each player has two options. The "cooperator" contributes $c > 0$ to a common pool, and the "defector" contributes nothing. The total contribution is multiplied by a factor of $r > 1$ and then shared equally among all ($n$) group members. A cooperator will thus pay a net cost $\sigma = c(1 - r/n)$ through its own contribution. If all cooperate, the group yields the optimal benefit $c(r-1)$ for each; if all defect, the group does nothing. To adhere to the spirit of the tragedy of the commons, we hereafter assume that $r < n$ holds, in which case a defecting player can improve its payoff by $\sigma > 0$, whatever the coplayers do, and the defectors dominate the cooperators. To observe the robustness for stochastic effects, we also consider an individual-based simulation with a pairwise comparison process [36,37]. See the electronic supplementary material (ESM) for these details.

**(a)** *Costly punishment*

We then introduce a third strategy, "punisher", which contributes $c$, and moreover, punishes the defectors. Punishing incurs a cost $\gamma > 0$ per defector to the punisher and imposes a fine $\beta > 0$ per punisher on the defector. We denote by $x$, $y$, and $z$ the frequencies of the cooperator (C), defector (D), and punisher (P), respectively. Thus, $x, y, z \geq 0$ and $x + y + z = 1$. Given the expected payoffs $P_S$ for the three strategies ($S$ = C, D, and P), the replicator system is written by

$$\dot{x} = x(P_C - \bar{P}), \quad \dot{y} = y(P_D - \bar{P}), \quad \dot{z} = z(P_P - \bar{P}), \tag{2.1}$$

where $\bar{P} := xP_C + yP_D + zP_P$ describes the average payoff in the entire population. Three homogeneous states ($x=1$, $y=1$, and $z=1$) are equilibria. Indeed,

$$P_C = \frac{rc}{n}(n-1)(x+z) - \sigma, \tag{2.2a}$$

$$P_D = \frac{rc}{n}(n-1)(x+z) - \beta(n-1)z, \tag{2.2b}$$

$$P_P = \frac{rc}{n}(n-1)(x+z) - \sigma - \gamma(n-1)y. \tag{2.2c}$$

Here the common first term denotes the benefit that resulted from the expected $(n-1)(x+z)$ contributors among the $(n-1)$ coplayers, and $\beta(n-1)z$ and $\gamma(n-1)y$ give the expected fine on a defector and expected cost to a punisher, respectively.

First, consider only the defectors and punishers (figure 1). Thus, $y + z = 1$, and the replicator system reduces to $\dot{z} = z(1-z)(P_P - P_D)$. Solving $P_P = P_D$ results in that, if the interior equilibrium R between the two strategies exists, it is uniquely determined by

$$z = 1 - \frac{(n-1)\beta - \sigma}{(n-1)(\beta + \gamma)}. \tag{2.3}$$

The point R is unstable. If the fine is much smaller: $\beta < \sigma/(n-1) =: \beta_0$, punishment has no effect on defection dominance, or otherwise, R appears and the dynamics turns into bistable [33,34]: R separates the state space into basins of attraction of the different homogeneous states for both the defector and excluder. The smaller $\gamma$ or larger $\beta$, the more the coordinate of R shifts to the defector end: the more relaxed the initial condition required to establish a punisher population (figure 1a). Note that a rare punisher is incapable of invading a defector population because the resident defectors, almost all unpunished, earn 0 on average, and the rare punisher does $-\sigma - \gamma(n-1) < 0$.

Next, consider all of the cooperators, defectors, and punishers (figure 1*b*). Without defectors, no punishing cost arises. Thus, no natural selection occurs between the cooperators and punishers, and the edge between the cooperators and punishers ($x + z = 1$) consists of fixed points. A segment consisting of these fixed points with $z > \beta_0/\beta$ is stable against the invasion of rare defectors, and the other segment not so [33,34]. Therefore, this stable segment appears on the edge PC if and only if the edge PD is bistable. We denote by $K_0$ the boundary point with $z = \beta_0/\beta$. There can thus be two attractors: the vertex D and segment $PK_0$. The smaller $\gamma$ or larger $\beta$, the broader the basin of attraction for the mixture states of the contributors. That is, the higher the punishment efficiency, the more relaxed the initial condition required to establish a cooperative state. This may collaborate with evidence from recent public-good experiments [38–40], which suggest the positive effects of increasing the punishment efficiency on average cooperation.

However, the stability of $PK_0$ is not robust for small perturbations of the population. Since $P_P < P_C$ holds in the interior space, an interior trajectory eventually converges to the boundary, and $d(z/x)/dt = (z/x)(P_P - P_C) < 0$: the frequency ratio of the punishers to cooperators decreases over time. Thus, if rare defectors are introduced, for example by mutation or immigration, into a stable population of the two types of contributors, the punishers will gradually decline for each elimination of the defectors. Such small perturbations push the population into an unstable regime around $K_0C$, where the defectors can invade the population and then take it over. See figure S1 of ESM and also [26] for individual-based simulations.

**(b)** *Social exclusion*

We turn next to social exclusion. The third strategy is now replaced with the excluder (E) that contributes *c* and also tries to exclude defectors from sharing benefits at a cost to itself of $\bar{\gamma} > 0$ per defector. The multiplied contribution is shared equally among the remaining members in the group. We assume that an excluder succeeds in excluding a defector with the probability $\bar{\beta}$ and that the excluded defector earns nothing. For simplicity, we conservatively assume that the total sanctioning cost for an excluder is given by $\bar{\gamma}$ times the number of defectors in a group, whatever others do.

We focus on perfect exclusion with $\bar{\beta}=1$: exclusion never fails. Under this condition, however, we can analyse the nature of social exclusion considered for cooperation. Indeed, we formalise the expected payoffs, as follows:

$$P_C = c(r-1) - (1-z)^{n-1} \frac{rc}{n}(n-1)\frac{y}{1-z}, \tag{2.4a}$$

$$P_D = (1-z)^{n-1} \frac{rc}{n}(n-1)\frac{x}{1-z}, \tag{2.4b}$$

$$P_E = c(r-1) - \bar{\gamma}(n-1)y. \tag{2.4c}$$

Equation (2.4c) describes that the excluder can constantly receive the group optimum $c(r-1)$ at the exclusion cost expected as $\bar{\gamma}(n-1)y$. In equations (2.4a) and (2.4b), $(1-z)^{n-1}$ denotes the probability that we find no excluder in the ($n-1$) coplayers, and if so, $(n-1)y/(1-z)$ and $(n-1)x/(1-z)$ give the expected numbers of the defectors and cooperators, respectively, among the coplayers. Hence, the second term of equations (2.4a) specifies an expected benefit that could have occurred without freeriding, and equation (2.4b) describes an expected amount that a defector has nibbled from the group benefit, in the group with no excluder. The expected payoffs for any $\bar{\beta}$ are formalised in ESM.

First, the dynamics between the excluders and defectors can only exhibit bi-stability or excluder dominance for $\bar{\beta}=1$ (figure 2a). Considering that $P_D = 0$ holds for whatever the fraction of excluders, solving $P_E = 0$ gives that, if the interior equilibrium R exists, it is uniquely determined by

$$z = 1 - \frac{(r-1)c}{(n-1)\bar{\gamma}}. \tag{2.5}$$

The point R is unstable. As before, for larger values of $\bar{\gamma}$, the dynamics between the two strategies have been bistable. The smaller the value of $\bar{\gamma}$, the larger the basin of attraction to the vertex E. In contrast to costly punishment, an excluder population can evolve, irrespective of the initial condition, for sufficiently small values of $\bar{\gamma}$. When decreasing $\bar{\gamma}$ beyond a threshold value, R exits at the vertex D, and thus, the current dynamics of bi-stability turns into excluder dominance. From substituting $z = 0$ into equation (2.5), the threshold value is calculated as $\bar{\gamma}_0 = (r-1)c/(n-1)$. We note that the dynamics exhibit defector dominance no matter what $\bar{\gamma}$, if $\bar{\beta}$ is smaller than $z_0$, which is from solving $(1-\bar{\beta})^{n-1} rc(n-1)/n > c(r-1)$: the unexcluded rare defector is better off than the resident excluders.

Next, consider all three strategies (figure 2b). Solving $P_C = P_D$ results in

$$z = 1 - \left(\frac{n(r-1)}{r(n-1)}\right)^{\frac{1}{n-1}} =: z_0. \qquad (2.6)$$

By the assumption $r < n$, we have $0 < z_0 < 1$. Let us denote by $K_0$ a point at which this line connects to the edge EC ($x + y = 1$). This edge consists of fixed points, each of which corresponds to a mixed state of the excluders and cooperators. These fixed points on the segment $EK_0$ ($z > z_0$) are stable, and those on the segment $K_0C$ are unstable. Similarly, solving $P_E = P_C$ gives

$$z = 1 - \left(\frac{n\gamma}{rc}\right)^{\frac{1}{n-2}} =: z_1. \qquad (2.7)$$

We denote by $K_1$ a point at which the line $z = z_1$ connects to EC. These two lines are parallel, and thus, there is no generic interior equilibrium.

Importantly, the time derivative of $z/x$ is positive in the interior region with $z < z_1$. Therefore, the dynamics around the segment $K_1K_0$ are found to be the opposite of costly punishment, if $z_1 > z_0$ (or otherwise, $K_1K_0$ has been unstable against rare defectors). In this case, introducing rare defectors results in that, for each elimination of the defectors, the excluders will gradually rise along $K_1K_0$ yet fall along the segment $EK_1$. Consequently, with such small perturbations, the population can remain attracted to the vicinity of $K_1$, not converging to D. Moreover, if $\bar{\gamma} < \bar{\gamma}_0$, the excluders dominate the defectors, and thus, all interior trajectories converge to the segment EK$_0$, which appears globally stable (figure 2*b*). This result remains robust for the intermediate exclusion probability (figure 3). See figures S2 and S3 of ESM for individual-based simulations.

## 3. Discussion

Our results regarding social exclusion show that it can be a powerful incentive and appears in stark contrast to costly punishment. What is the logic behind this outcome? First, it is a fact that the exclusion of defectors can decrease the number of beneficiaries, especially when it does not affect the contributions, thereby increasing the share of the group benefit. Therefore, in a mixed group of excluders and defectors, the excluder's net payoff can become higher than the excluded defector's payoff, which is nothing, especially if the cost to exclude is sufficiently low. If social exclusion is capable of 100% rejection at a cheap cost, it can thus emerge in a sea of defectors and dominate them. In our model, self-serving punishment can emerge even when freeriding is initially prevalent by allowing high net benefits from the self-serving action.

Moreover, we find that an increase in the fraction of excluders produces a higher probability of an additional increase in the excluder's payoff. This effect can yield the well-known Simpson's paradox (e.g., [41]): the excluders can obtain a higher average payoff than the cooperators, despite the fact that the cooperators always do better than the excluders for any mixed group of the cooperators, defectors, and excluders. Hence, in the presence of defectors, the replicator dynamics often favour the excluders at the expense of the cooperators. Significantly, if a player may occasionally mutate to a defector, social exclusion is more likely than costly punishment to sustain a cooperative state in which all contribute. In our model, a globally stable, cooperative regime can be sustained when solving the second-order freerider problem by allowing mutation to freeriders.

Sanctioning the second-order freeriders has also often been considered for preventing their proliferation [3,29,34,36], although such second-order sanction appears rare in experimental settings [42]. And, allowing for our simple model, it is obvious that in the presence of defectors and cooperators, a second-order punisher that also punishes the cooperators is worse off than the existing punisher, and thus, does not affect defector dominance as in our main model. However, given that excluding more coplayers can cause an additional increase in the share of the group benefit, it is worth exploring whether the second-order excluder that also excludes the cooperators is more powerful than the excluder. Interestingly, our preliminary individual-based investigation often finds that second-order excluders are undermined by the excluders and cooperators, which forms a stable coexistence (figures S4 of EMS): second-order exclusion can be redundant.

A fundamental assumption of the model is that defection can be detected with no or little cost. This assumption appears most applicable to local public goods and team production settings in which the coworker's contribution can be easily monitored. However, if the monitoring of co-players for defection imposes a certain cost on the excluders, the cooperators dominate the excluders, and the exclusion-based full cooperation is no longer stable. A typical example is found in a potluck party that will often rotate so that every member takes charge of the party by rotation. This rotation system can promote the equal sharing of the hosting cost; or otherwise, no one would take turn playing host.

We assessed by extensive numerical investigations the robustness of our results with respect to the following variants (figures S5 and S6 of EMS). First, we considered a different group size $n$ [3,43], In costly punishment, the stable segment $PK_0$ expands with $n$, yet our main results were unaffected: with small perturbations, the population eventually converges to a non-cooperative state in which all freeride. In social exclusion, our results remain qualitatively robust with smaller and larger sizes ($n = 4$ and $n = 10$), but the limit exclusion cost $\bar{\gamma}$ becomes more restricted as $n$ increases. Next, we considered a situation in which a punisher or excluder can choose the number of defectors they sanction. For simplicity, here we assume that each of them sanctions only one [22,44], who is selected randomly from all defectors in the group. Our results remain unaffected, except that social exclusion becomes incapable of emerging in a defector population, in which the payoff of a rare excluder is only given by $rc/(n-1) - c - \bar{\gamma} < 0$. To bring forth the possibility of an emergence, a rare excluder is required to exclude more than $n - rc/(c + \bar{\gamma})$ defectors.

Our results spur new questions about earlier studies on the evolution of cooperation with punishment. A fascinating extension is to the social structures through which individuals interact. To date, a large body of work on cooperation has looked at how costly punishment can propagate throughout a social network [45–47]: for example, the interplay of costly punishment and reputation can promote cooperation [48]; strict-and-severe punishment and cooperation can jointly evolve with continuously varying strategies [49]; and evolution can favour anti-social punishment that targets cooperators [50]. Our results show that social exclusion as considered is so simple, yet extremely powerful. That is, even intuitively applying it to previous studies can help us much in understanding how humans and non-humans have been incentivized to exclude freeriders.

To resist the exclusion, it is likely that conditional cooperators capable of detecting ostracism (e.g., [8]) evolve. This would then raise the comprehensive cost of exclusion to the excluders because of more difficulties of finding and less opportunities of excluding freeriders. This situation can then result in driving an arms race of the exclusion technique and exclusion detection system. An extensive investigation for understanding joint evolution of these systems is for future work.

**Acknowledgements**

We thank Karl Sigmund and Voltaire Cang for their comments and suggestions. This study was enabled by financial support by the FWF (Austrian Science Fund) to Ulf Dieckmann at IIASA (TECT I-106 G11), and was also supported by grant RFP-12-21 from the Foundational Questions in Evolutionary Biology Fund.

**Figure captions**

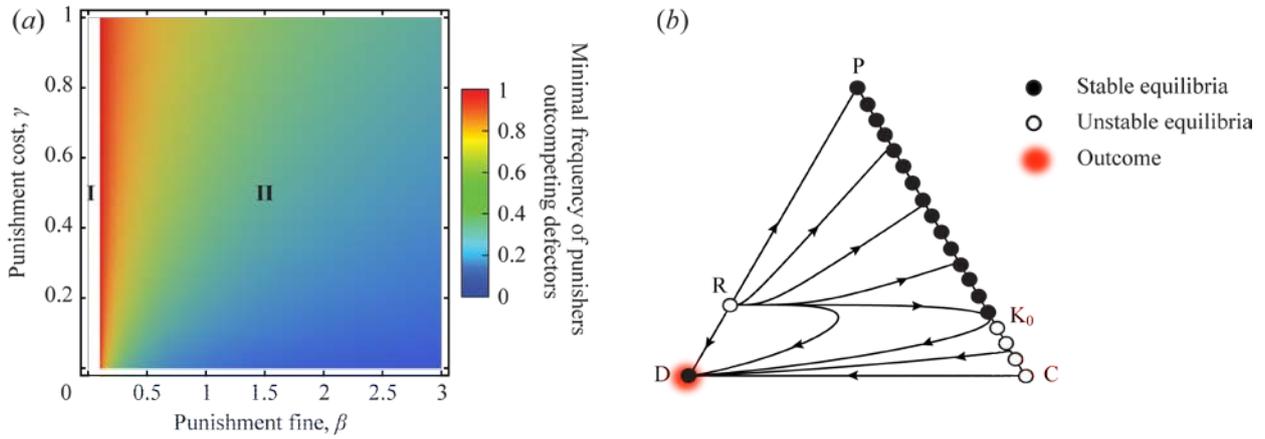

Figure 1. Effects of punishing freeriders. (*a*) Between the punishers and freeriders. **I,** If $\beta$ is smaller than a threshold value $\beta_0 = \sigma/(n-1)$, where $\sigma = c(1-r/n)$ describes a net cost for the single contributor, the defectors dominate. **II,** If $\beta$ is greater than $\beta_0$, punishing leads to bistable competition between the two strategies. With increasing $\beta$ or decreasing $\gamma$, the minimal frequency of the punishers outcompeting the defectors decreases. However, the excluders cannot dominate the defectors for finitely large values of $\beta$. Parameters: group size $n=5$, multiplication factor $r=3$, and contribution cost $c=1$. (*b*) In the presence of second-order freeriders. The triangle represents the state space, $\Delta = \{(x,y,z): x,y,z \geq 0,\ x+y+z=1\}$, where *x*, *y*, and *z* are the frequencies of the cooperators, defectors, and punishers, respectively. The vertices, C, D, and P, correspond to the three homogeneous states in which all are the cooperators ($x=1$), defectors ($y=1$), or punishers ($z=1$). The edge PC consists of a continuum of equilibria. The defectors dominate the cooperators. Here we specifically assume $\beta = 0.5$ and $\gamma = 0.03$, which result in an unstable equilibrium R within PD and the segmentation of PC into stable part $PK_0$ and unstable part $K_0C$. The interior of $\Delta$ is separated into the basins of attraction of D and $PK_0$. In fact, given the occasional mutation to a defector, the population's state must leave $PK_0$ and then enter the neighborhood of the unstable segment $K_0C$ because $P_P > P_C$ holds over the interior space. The population eventually converges to D.

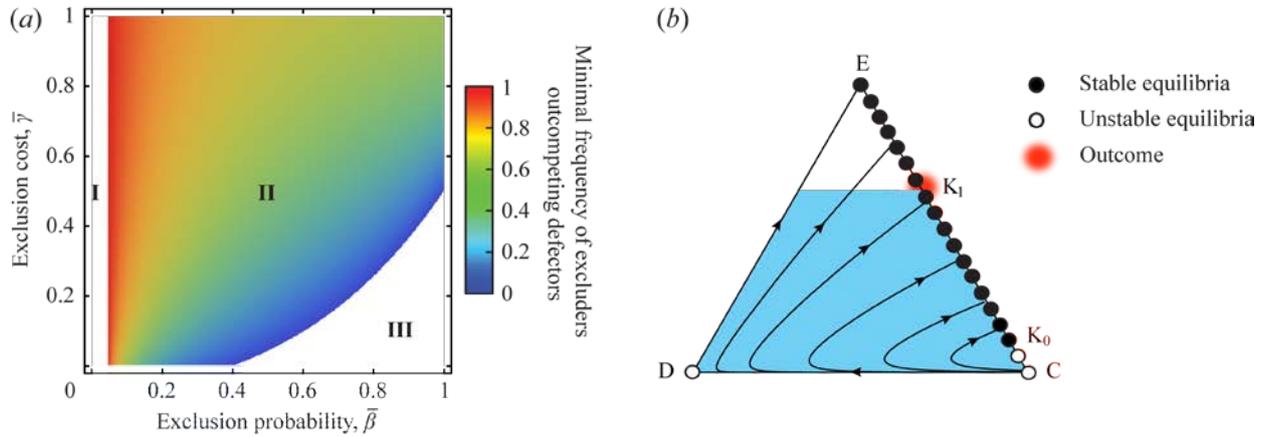

Figure 2. Effects of excluding freeriders. (*a*) Between the excluders and freeriders. **I,** If $\bar{\beta}$ is smaller than a threshold value $z_0$, the defectors dominate. **II,** If $\bar{\beta}$ is greater than $z_0$, exclusion leads to bistable competition between the two strategies. With increasing $\bar{\beta}$ or decreasing $\bar{\gamma}$, the minimal frequency of the excluders outcompeting the defectors decreases. **III,** If $\bar{\beta}$ and $\bar{\gamma}$ are sufficiently high and low, the excluders dominate. The parameters are as in figure 1*a*. (*b*) In the presence of second-order freeriders. The triangle Δ is as in figure 1*b*, except that *z* denotes the excluder frequency and the vertex E corresponds to its homogeneous state. Similarly, the edge EC consists of a continuum of equilibria. Here we specifically assume $\bar{\beta}=1$ and $\bar{\gamma}=0.03$. EC is separated into stable and unstable segments. The coloured area in the interior of Δ is the region in which $P_E > P_C$ holds. In fact, given the occasional mutation to a defector, the population's state must converge to the vicinity of the point $K_1$, because the advantage of the excluders over the cooperators becomes broken when the population's state goes up beyond $K_1$.

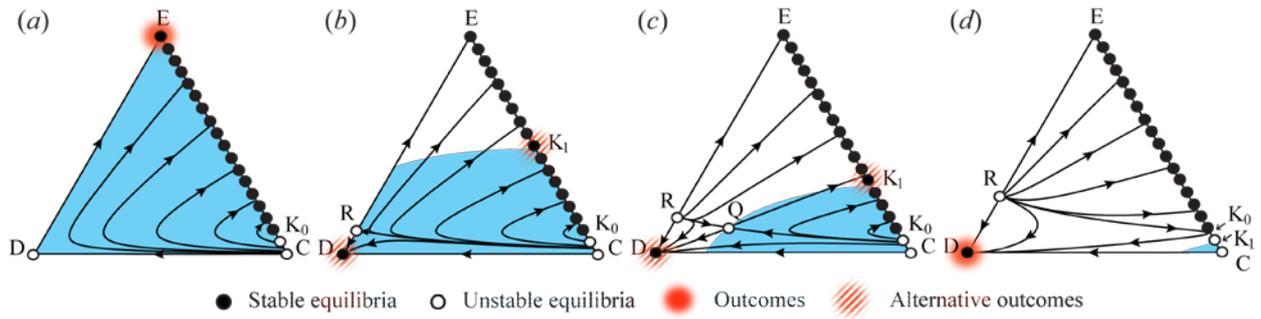

Figure 3. Effects of intermediate social exclusion in the presence of second-order freeriders. The parameters and triangles are as in figure 1, except that $\bar{\beta} = 0.5$ and $\bar{\gamma} = 0.03$ (*a*), 0.13 (*b*), 0.18 (*c*), or 0.28 (*d*). EC is separated into stable and unstable segments. The coloured area is the interior region in which $P_E > P_C$ holds. (*a*) The dynamics of ED are unidirectional to E. All interior trajectories converge onto the stable segment $EK_0$. Moreover, occasionally mutating to a defector leads to upgrading E to a global attractor. (*b-d*) An unstable equilibrium R appears on ED. The interior space is separated into the basins of attraction of D and $EK_0$. R is a saddle (*b*) or source (*c* and *d*). In (*c*) especially, the interior space has a saddle point Q. Given the mutant defectors, the population's state around $EK_0$ will gradually move to $K_1$ (*b* and *c*), or to the unstable segment $K_0C$ (*d*). The last case is followed by a convergence toward D.

1    Electronic supplementary material (ESM)

2    This includes: Materials and methods, and Supplementary figures, S1–S6

3    **Materials and methods**

4    We first determine the strategy's payoffs in public good games with social exclusion, then
5    show details of individual-based simulations for assessing the robustness with respect to
6    stochastic evolutionary game dynamics.

7    **Payoffs for social exclusion:** We consider the replicator dynamics for the cooperator (C),
8    defector (D), and excluder (E), with frequencies of $x$, $y$, and $z$, respectively. Thus, $x, y, z \geq 0$
9    and $x + y + z = 1$. We denote the expected payoff values for the three strategies by $P_S$, with $S =$
10   C, D, and E, respectively. The replicator system is given by
11   $\dot{x} = x(P_C - \bar{P}), \ \dot{y} = y(P_D - \bar{P}), \ \dot{z} = z(P_E - \bar{P}),$
12   where $\bar{P} := xP_C + yP_D + zP_E$ describes the average payoff in the entire population. We denote by
13   $X$, $Y$, and $Z$ the number of the cooperators, defectors, and excluders, respectively, among the
14   $(n-1)$ coplayers around a focal player. Then, if $W$ of the $Y$ defectors have not been excluded
15   by every excluder, the expected payoff for each strategy is given by

16   $$P_S = \sum_{X=0}^{n-1} \sum_{Y=0}^{n-1-X} \sum_{W=0}^{Y} \pi_S p_S.$$  (S1)

17   In equation (S1), $p_S$ denotes the payoff for the focal player who follows the strategy $S$ among
18   the $(n-1)$ coplayers with a configuration of $\{X, Y, Z, W\}$, and $\pi_S$ denotes the probability to
19   find the specified coplayers. Using a function $\alpha(Z)$ that denotes the probability that all of the
20   $Z$ excluders fail to exclude a targeted defector, we have

21   $$p_C = \frac{rc(X + Z + 1)}{X + W + Z + 1} - c,$$  (S2)

22   $$p_D = \alpha(Z) \frac{rc(X + Z)}{X + W + Z + 1},$$  (S3)

23   $$p_E = p_C - \bar{\gamma} Y,$$  (S4)

24  $$\pi_{\text{C}} = \pi_{\text{D}} = \binom{n-1}{X,Y,Z} x^X y^Y z^Z \binom{Y}{W} \alpha(Z)^W [1-\alpha(Z)]^{Y-W}, \tag{S5}$$

25  $$\pi_{\text{E}} = \binom{n-1}{X,Y,Z} x^X y^Y z^Z \binom{Y}{W} \alpha(Z+1)^W [1-\alpha(Z+1)]^{Y-W}. \tag{S6}$$

26  In equations (S5) and (S6), $\binom{n-1}{X,Y,Z}$ and $\binom{Y}{W}$ represent the multinomial and binomial

27  coefficients. Thus, $\binom{n-1}{X,Y,Z} x^X y^Y z^Z$ describes the probability of finding the ($n-1$) coplayers

28  with $X$ cooperators, $Y$ defectors, and $Z$ excluders, and $\binom{Y}{W} \alpha(Z)^W [1-\alpha(Z)]^{Y-W}$ describes the

29  probability that $W$ of the $Y$ defectors have not been excluded. In the paper, we assume

30  $\alpha(Z) = (1-\bar{\beta})^Z$, where $\bar{\beta}$ is the exclusion probability: an excluder succeeds in excluding a

31  defector.

32  **Individual-based simulation**: Here, we consider a finitely large, well-mixed population with

33  $M$ interacting individuals. For the dynamic analysis, instead of the replicator system [35], we

34  implement a pairwise comparison process among finite individuals [36,37], which is based on

35  preferentially imitating strategies of more successful individuals. We assume that the

36  individual strategies are updated asynchronously as follows. First, an individual $i$ is selected at

37  random and then earns its "average" payoff $p_i$ after engaging in $T$ games with coplayers

38  randomly selected in each case. Second, the focal individual $i$ faces a model individual $j$ who is

39  drawn at random, with its average payoff $p_j$ that is calculated throughout independent $T$

40  games. If $p_i \geq p_j$, no update occurs; or otherwise, $i$ will adopt $j$'s strategy, with the probability

41  given by

42  $$\theta_{i \to j} = \frac{1}{1+\exp(-K(p_j - p_i))},$$

43  where $K$ denotes the selection strength. Finally, the focal individual $i$ can mutate and turn into a

44  cooperator, defector, or punisher (or excluder) with probabilities $\mu_C$, $\mu_D$, $\mu_P$ (or $\mu_E$). Our

45  numerical results demonstrated in figures S1–S6 are robust with respect to changes in the

46  parameter values of $M$, $\mu_C$, $\mu_D$, $\mu_P$, $\mu_E$, and $K$.

**Supplementary figures**

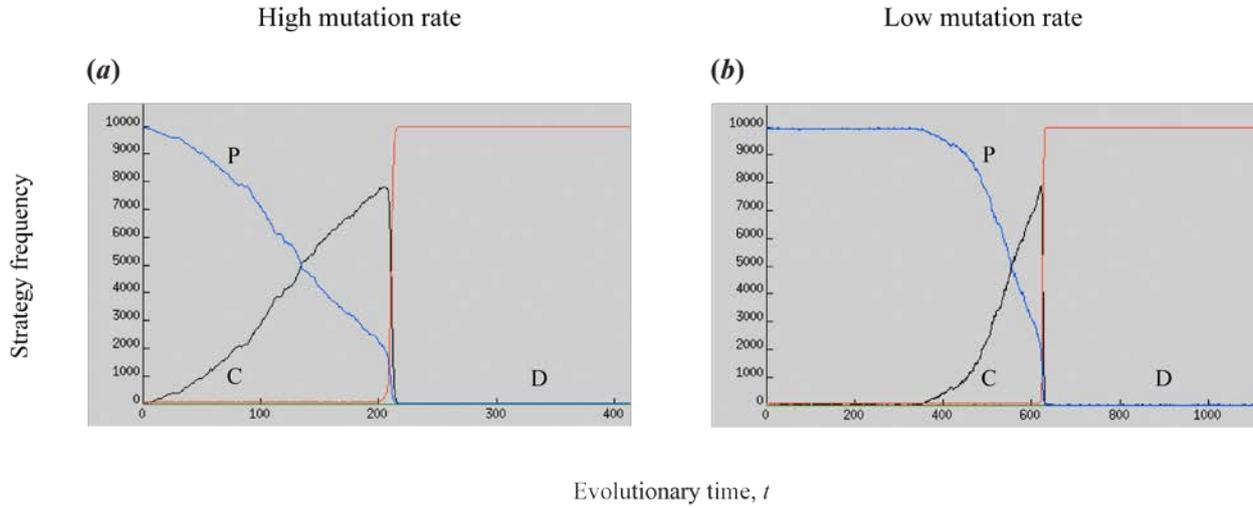

Figure S1. Individual-based simulation for public good games with costly punishment. We began with a 100%-punisher population to observe its stability. First, because the punishing of mutant defectors is costly, the former major punishers (blue) will gradually be replaced by the initially minor cooperators (namely, second-order freeriders, black). Next, when a critical fraction of punishers is lost, the mutant defectors (red) succeed in invading the population and then quickly prevail. The parameters are as in figure 1b: group size $n=5$, multiplication factor $r=3$, contribution cost $c=1$, punishment cost $\beta=0.5$, and punishment fine $\gamma=0.03$. The defectors dominate the cooperators, and the excluders and defectors are under bistable competition. Other parameters are as the population size $M=10^4$, sample game count $T=50$, selection strength $K=200$, mutation rate to D $\mu_D=5\times 10^{-3}$, mutation rates to C and P $\mu_C=\mu_P=10^{-5}$ (low mutation rate) or $\mu_C=\mu_P=10^{-3}$ (high mutation rate), and the unit of evolutionary time $t$ describes $10^4$ times the iteration of the update sequence.

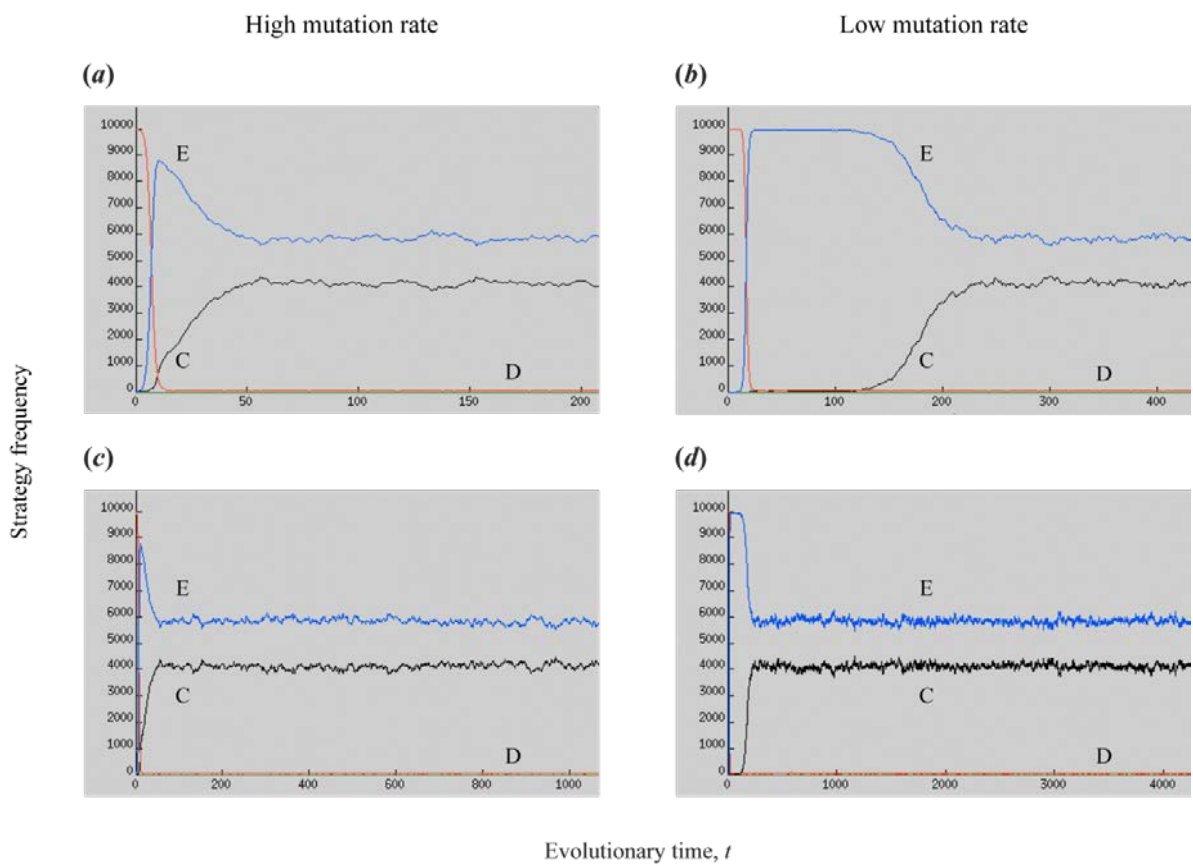

Figure S2. Individual-based simulation for public good games with perfect social exclusion. The parameters are as in figure 2b: $n=5$, $r=3$, $c=1$, exclusion probability $\bar{\beta}=1$, and exclusion cost $\bar{\gamma}=0.03$. We began with a 100%-punisher population to observe the establishment of a cooperative state. Whether the minimal mutation rate is high ($10^{-3}$) or low ($10^{-5}$), the former major defectors (red) will soon be replaced by the initially minor excluders (blue), whose part will then be gradually replaced by the cooperators (black). The population eventually converges to a certain mixture state of the contributors without a second-order freerider problem. The final state has been indicated by point $K_1$ in figures 2b. The simulation parameters are as in figure S1.

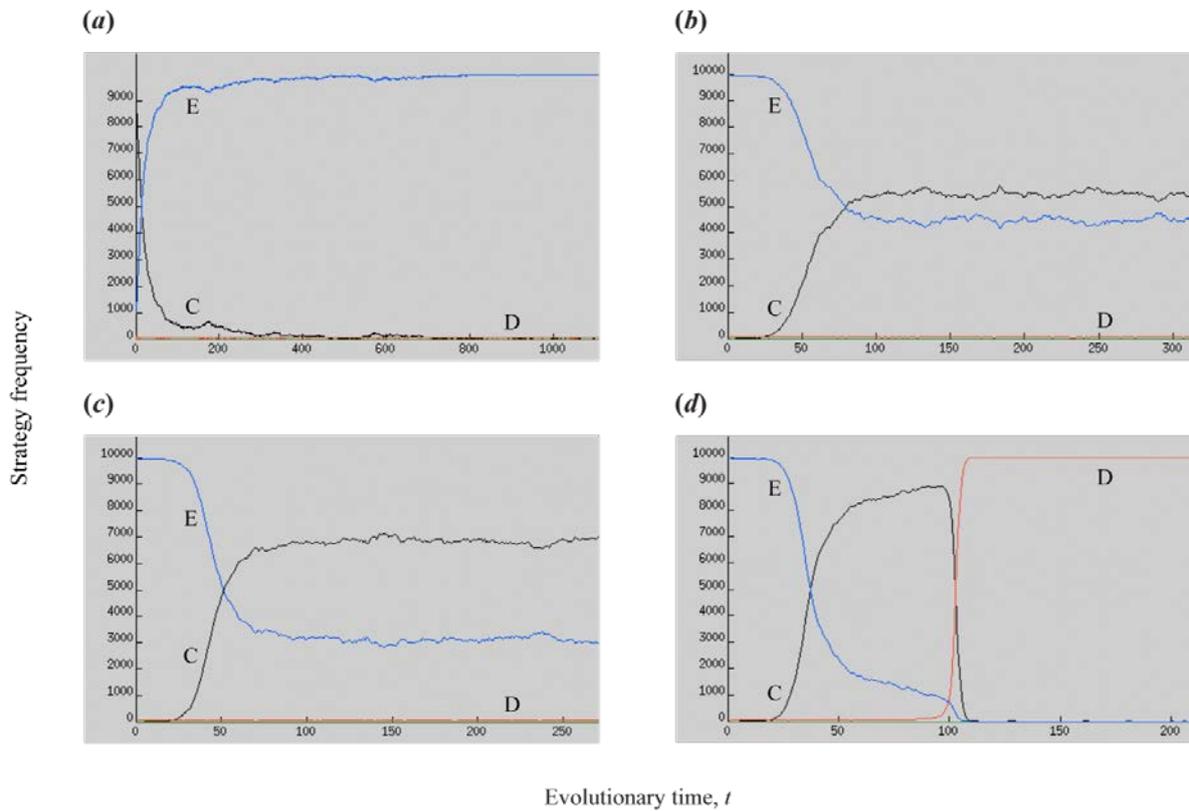

Figure S3. Individual-based simulation for public good games with intermediate social exclusion. The parameters are as in figure 3: $n = 5$, $r = 3$, $c = 1$, and $\bar{\beta} = 0.5$. We began with different initial conditions, depending on the value of $\bar{\gamma}$: 90% cooperators and 10% excluders for $\bar{\gamma} = 0.03$ (*a*) and 100% excluders for $\bar{\gamma} = 0.13$ (*b*), 0.18 (*c*), or 0.28 (*d*). (*a*) The former major cooperators (black) will gradually be replaced by the initially minor excluders (blue), which then stably occupy the entire population (*b* and *c*). The initially minor cooperators will first replace part of the excluders, and the population will then converge to a certain mixture state, which has been indicated by the point $K_1$ in figures 3*b* and 3*c*, respectively (*d*). As in (*b* and *c*), the cooperators will gradually expand. When a critical fraction of the excluders is lost (the point $K_0$), the mutant defectors (black) succeed in invading the population and will then quickly prevail to 100%. The simulation parameters are as in figure S1 with the low mutation rate.

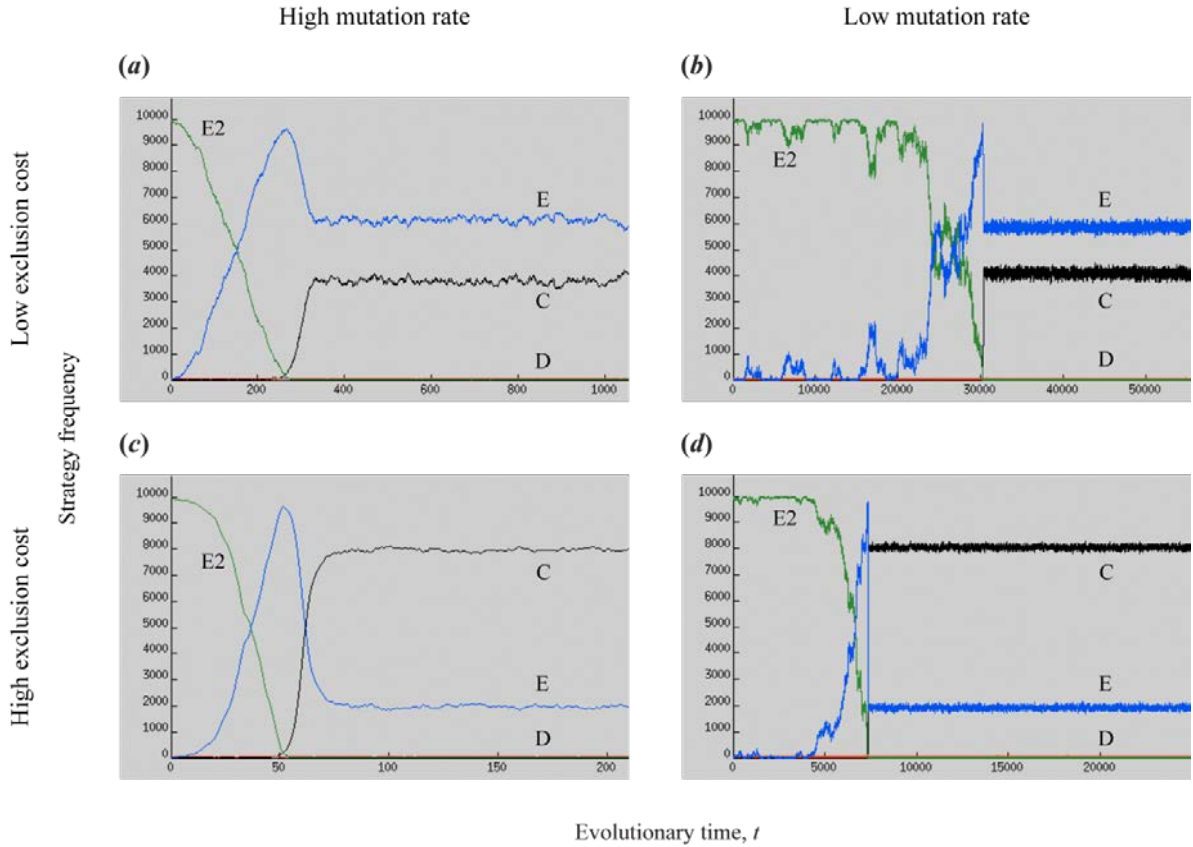

Figure S4. Individual-based simulation for public good games with second-order social exclusion. The parameters are as in figure 2b, except that $\bar{\gamma}=0.03$ (low exclusion cost) or $\bar{\gamma}=0.28$ (high exclusion cost). We began with the initial condition: 100% second-order excluders (green) who in the presence of the defectors, also exclude the cooperators, as well as the defectors (with the same cost and probability). The initial residents will first be replaced with the excluders (blue), and then are partially invaded by the cooperators (black): the population will converge to a certain mixture state of the contributors, whether with a high or low exclusion cost. The simulation parameters are as in figure S1.

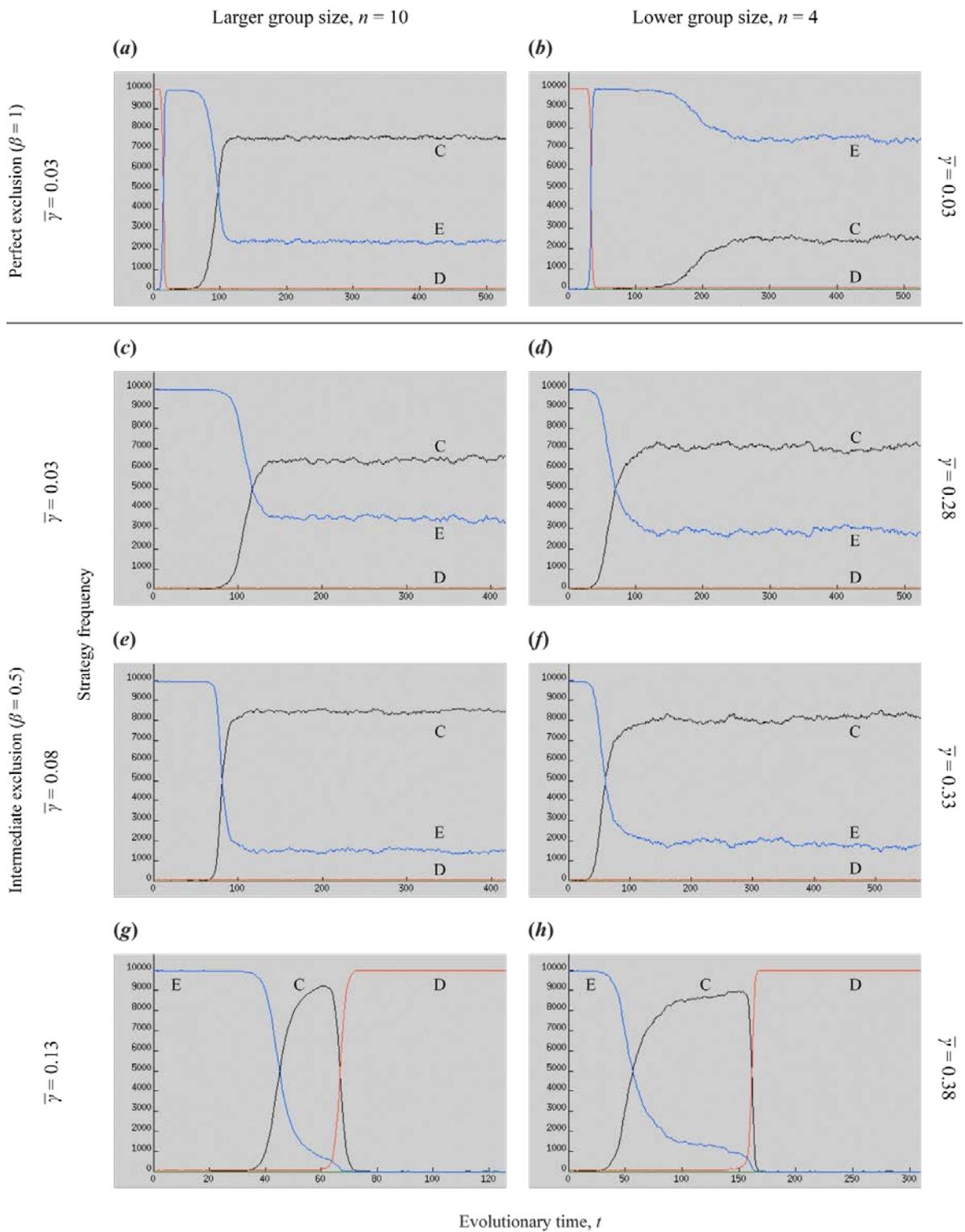

Figure S5. Effect of different group sizes. The parameters are as in figure 2b, for perfect exclusion (*a*) and (*b*), and in figure 3, for intermediate exclusion (*c–h*). The initial conditions are 100% second-order excluders in (*a*) and (*b*) and 100% excluders in (*c–h*).

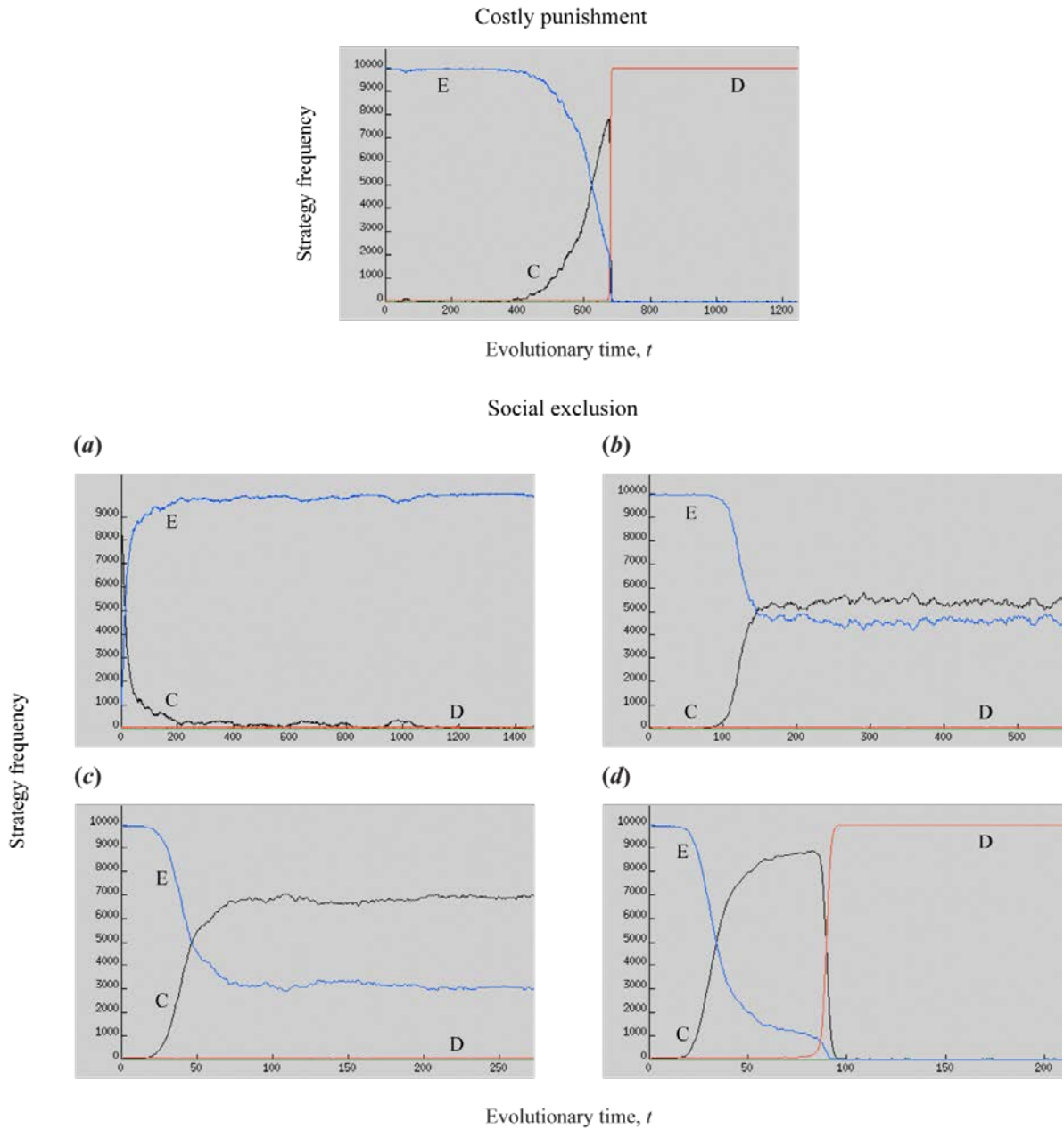

Figure S6. Effect of options to choose the number of sanctioned defectors. The model and simulation parameters, and initial conditions are as in figure S1, for costly punishment (top), and in figure S3, for intermediate exclusion (middle and bottom, *a–d*). Here we assume that a punisher or excluder is willing to sanction only one defector selected at random from all defectors in the group. The results are almost same as in figures S1 and S3.